\documentclass[aps,prl,reprint,superscriptaddress]{revtex4-2}

\usepackage{amsmath,amssymb}
\usepackage{graphicx}
\usepackage{bm}
\usepackage{dcolumn}
\usepackage{url}

\begin{document}

\title{Parametric Enhancement of an Optical Spring in a Dark Signal-Recycling Cavity}

\author{Kaido Suzuki}
\email{ksuzuki@gw.phys.sci.isct.ac.jp}
\affiliation{Department of Physics, Institute of Science Tokyo, 2-12-1 Oh-okayama, Meguro, Tokyo 152-8551, Japan}
\affiliation{School of Physics, Mathematics and Computing, The University of Western Australia, 35 Stirling Highway, Perth, WA 6009, Australia}
\author{Ryo Iden}
\affiliation{Department of Physics, Institute of Science Tokyo, 2-12-1 Oh-okayama, Meguro, Tokyo 152-8551, Japan}
\author{Ken-ichi Harada}
\affiliation{Nanofiber Quantum Technologies, Inc., 1-22-3 Nishiwaseda, Shinjuku, Tokyo 169-0051, Japan}
\author{Kentaro Somiya}
\affiliation{Department of Physics, Institute of Science Tokyo, 2-12-1 Oh-okayama, Meguro, Tokyo 152-8551, Japan}

\date{\today}

\begin{abstract}
Optical springs tailor the response of precision sensors. Parametric amplification of the signal field stiffens the spring, extending the precision to higher frequencies. A Michelson interferometer operated at a dark fringe is the most direct way to separate the signal from the carrier for amplification. We experimentally demonstrate such quantum manipulation of the optical spring by placing an optical parametric amplifier in a cavity at the dark port, showing a significant increase of the optical-spring constant with the parametric gain within a factor of four below the oscillation threshold.
\end{abstract}

\maketitle

\section{Introduction}\label{sec:introduction}

Optical cavities enhance weak signals in precision measurements, from cavity optomechanical sensing~\cite{Aspelmeyer2014} and molecular spectroscopy~\cite{Gianfrani2024} to atomic magnetometry~\cite{Crepaz2015} and gravitational-wave detection~\cite{Abbott2016}. In most cavity-enhanced interferometers, however, the signal response is closely tied to the circulating carrier power, and raising that power eventually introduces thermal loading~\cite{Carmon2004}, optical distortion~\cite{Brooks2016}, and optomechanical instabilities~\cite{Kippenberg2005,Evans2015}. Amplifying the signal instead would lift this constraint.

A Michelson interferometer operated at a dark fringe, as is common in gravitational-wave detectors~\cite{Aasi2015}, naturally separates the signal and carrier fields; the carrier returns towards the symmetric port, whereas differential signal sidebands emerge from the antisymmetric port. A signal-recycling mirror (SRM) at this port forms a signal-recycling cavity, which stores the differential signal but is dark for the carrier. An optical parametric amplifier (OPA) inside this cavity enables phase-sensitive manipulation of the signal and quantum-noise fields while leaving the circulating carrier power unchanged. This approach has been explored theoretically in the form of intracavity squeezing to improve high-frequency sensitivity and overcome the conventional sensitivity--bandwidth trade-off~\cite{Korobko2019,Adya2020}. Like externally injected squeezing, already employed in a range of precision measurements~\cite{LIGO2011Squeezing,Horrom2012,Hoff2013}, however, intracavity squeezing improves sensitivity through quantum-noise reduction rather than signal enhancement. Complementary proposals exploit the optical spring, an effective restoring force produced by radiation-pressure feedback in a detuned interferometer~\cite{BuonannoChen2002}, which has been used to tailor mechanical dynamics and suppress noise~\cite{Corbitt2007Dilution,Eichenfield2009,Cullen2022}. Intracavity parametric amplification has been proposed either to strengthen the optical spring~\cite{Somiya2016,Korobko2018,Somiya2023} or to enhance radiation-pressure cooling of mechanical oscillators~\cite{Huang2009}.

Such enhancement has been demonstrated in single-cavity configurations, using either an effective Kerr interaction~\cite{Otabe2024} or an OPA~\cite{Zhang2023, Liu2025}. In these configurations, however, the nonlinear medium is exposed to the circulating carrier field, and absorption-induced photothermal effects can modify both the cavity response and the mechanical susceptibility, complicating control of the spring~\cite{Otabe2022}. Separating the signal field from the carrier allows the same enhancement without this thermal load.

Here we place an OPA in the detuned signal-recycling cavity of a Michelson interferometer and coherently control its phase together with the interferometer and cavity degrees of freedom. The optical spring provides an in situ probe of the intracavity signal response. Measurements over a range of cavity detunings show that activating the amplifier increases the peak optical-spring constant by 8.2$\pm$2.8\% without increasing the carrier power, with the parametric gain a factor of four below the oscillation threshold.

\section{Parametric enhancement of the optical spring}\label{sec:results}

We consider a Michelson interferometer operated at a dark fringe, with a phase-sensitive optical parametric amplifier inside the detuned signal-recycling cavity. Differential mirror motion generates signal sidebands that enter this cavity and return to the interferometer, where they beat with the carrier to produce a radiation-pressure force. The resulting feedback creates an optical spring, whose stiffness can be increased by amplifying the signal field.

\begin{figure}[htbp]
\centering
\begin{minipage}[t]{0.49\textwidth}
  \begin{minipage}[t]{0.49\linewidth}
    \textbf{(a)}\par\vspace{0.5mm}
    \includegraphics[width=\linewidth,trim={0 0.7 19 5},clip]{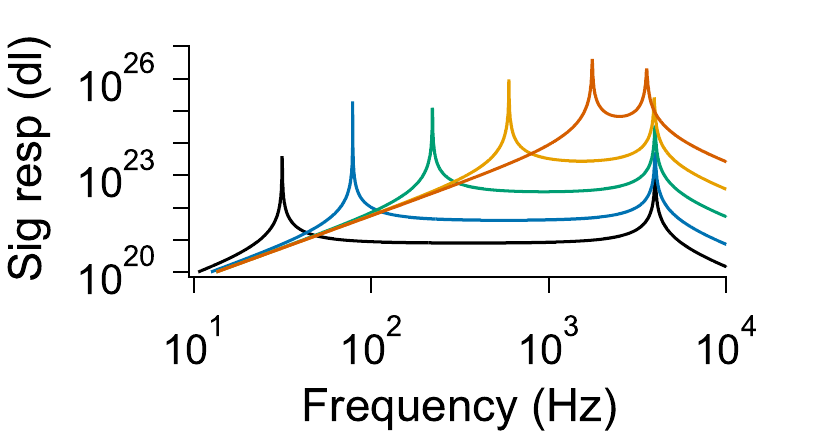}
  \end{minipage}%
  \hfill
  \begin{minipage}[t]{0.49\linewidth}
    \textbf{(b)}\par\vspace{0.5mm}
    \includegraphics[width=\linewidth,trim={0 0.7 19 5},clip]{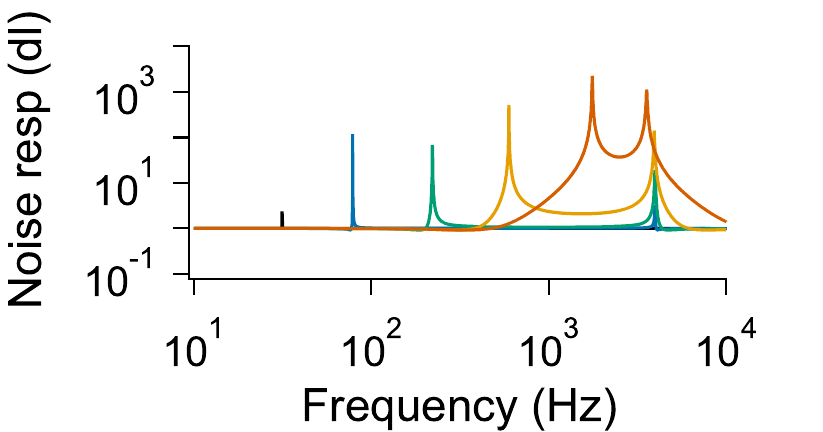}
  \end{minipage}
  \vspace{3mm}
  \textbf{(c)}\par\vspace{-1.2ex}
  \hspace*{0.05\linewidth}\includegraphics[width=0.7\linewidth,trim={0.7 0 12 2},clip]{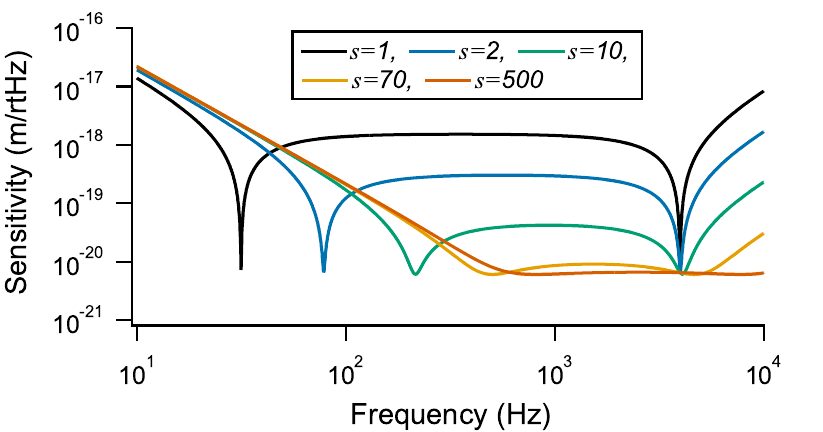}
\end{minipage}
\caption{\label{fig:active-enhancement}Parametric enhancement of an optical spring in a dark signal-recycling cavity. (a),(b) Signal and quantum-noise response functions, respectively; dl denotes dimensionless. (c) Quantum-noise-limited displacement sensitivity. The curves in (a)--(c) correspond to $s=1$, $2$, $10$, $70$ and $500$, with $s=1$ representing the passive interferometer. The detuning is adjusted with gain to keep the optical resonance frequency fixed. Optical losses are neglected.}
\end{figure}

Including the amplification-axis angle $\eta$, the low-frequency optical-spring frequency in the lossless model is
\begin{equation}
f_{\rm OS}^2 \simeq f_0^2\,\frac{\tfrac12\left[(s+s^{-1})\sin 2\phi-(s-s^{-1})\sin(2\phi+2\eta)\right]}{(r+r^{-1})-(s+s^{-1})\cos 2\phi},
\label{eq:fos}
\end{equation}
where $s$ is the OPA amplitude gain, with $s=1$ corresponding to the passive cavity, $r$ is the amplitude reflectivity of the SRM, $\phi$ is the cavity detuning phase, and $f_0^2 = 4\omega_0 P_{\rm BS}/[(2\pi)^2 m c^2]$ sets an overall frequency scale independent of detuning and gain, with $\omega_0$ the carrier angular frequency, $P_{\rm BS}$ the carrier power at the beam splitter and $m$ the test mass. The amplification-axis convention is $\bm{S}(s,\eta)=\bm{R}(\eta)\operatorname{diag}(s,s^{-1})\bm{R}(-\eta)$. The parametric oscillation threshold is reached when the denominator of Eq.~\eqref{eq:fos} vanishes. Operation below this threshold requires
\begin{equation}
(r+r^{-1})-(s+s^{-1})\cos 2\phi>0.
\label{eq:stability}
\end{equation}

Throughout Fig.~\ref{fig:active-enhancement}, $\eta$ is set to maximize the optical-spring frequency at each gain and detuning: $\phi+\eta=-\pi/4$ for $s>1$ and $+\pi/4$ for $s<1$. This optimization gives the symmetry under $s\leftrightarrow s^{-1}$. The optical resonance frequency is held fixed by setting
\begin{equation}
\phi(s)=\frac{1}{2}\arccos\left[\frac{2}{s+s^{-1}}\cos 2\phi_0\right].
\label{eq:phi_of_s}
\end{equation}
The calculations in Fig.~\ref{fig:active-enhancement} use the input--output model, notation and parameters of Ref.~\cite{Somiya2023}. The SRM power reflectivity is 99.95\%, the arm length is 1200 m, the carrier power at the beam splitter is 100 kW, the test mass is 1 kg, and the initial detuning is $\phi_0 = 5.74^\circ$, chosen to place the optical resonance frequency at 4 kHz. Internal optical loss and detection loss are set to zero, and the output is read out at a homodyne angle $\zeta = 0$.

The enhancement changes both the mechanical response and the quantum-noise-limited sensitivity. Increasing the parametric gain shifts the optical-spring resonance to higher frequency and enhances the signal response above the resonance (Fig.~\ref{fig:active-enhancement}(a)), while the noise response remains nearly unchanged until the gain becomes sufficiently large (Fig.~\ref{fig:active-enhancement}(b)). Consequently, the sensitivity dip associated with the optical spring shifts to higher frequency, before disappearing at high gain (Fig.~\ref{fig:active-enhancement}(c)). In this high-gain regime, the detuning approaches $\pi/4$, and the broadband improvement at high frequencies connects to cyclic-quadrature intracavity signal amplification~\cite{Suzuki2026}.

A proof-of-principle experiment is essential for this new technique. The manipulation of the optical spring with the intracavity OPA can be examined by measuring the optomechanical response function shown in Fig.~\ref{fig:active-enhancement}(a). By keeping the optical resonance frequency and circulating power fixed, a shift of the optical spring frequency can be observed with a moderate OPA gain.

\section{Apparatus and measurement}\label{sec:experiment}

\begin{figure}[htbp]
\centering
  \includegraphics[width=\linewidth,trim={5 4 6 4},clip]{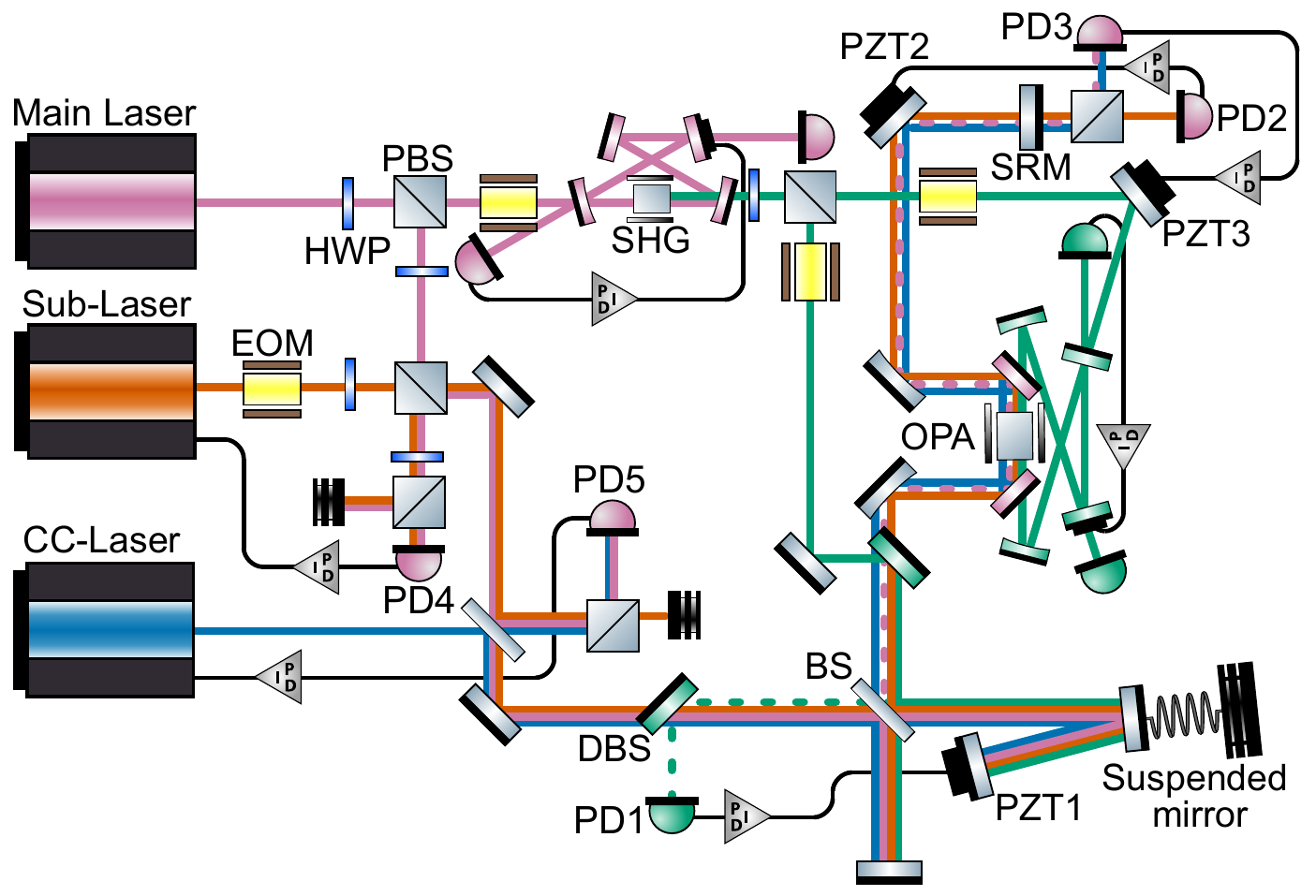}
\caption{\label{fig:experiment}Experimental setup. Michelson interferometer with an OPA inside the signal-recycling cavity and a resonant pump cavity. Auxiliary 532-nm, sub-carrier and coherent-control (CC) fields enable independent control of the interferometer, the signal-recycling cavity and the OPA phase, respectively. PZT1--3 actuate the corresponding optical path lengths. HWP, half-wave plate; EOM, electro-optic modulator; PBS, polarizing beam splitter; DBS, dichroic beam splitter; BS, beam splitter; SHG, second-harmonic generation; PD, photodetector; PZT, piezoelectric actuator.}
\end{figure}

The apparatus (Fig.~\ref{fig:experiment}) extends the signal-recycled Michelson interferometer in which an optical spring was previously observed without the intracavity amplifier~\cite{Suzuki2025}. Values quoted for the apparatus are nominal. A 1064-nm carrier illuminated the interferometer with $300\,\mathrm{mW}$ in each arm. The 200-mg folding mirror, suspended by a double-sided BeCu spiral spring, had a mechanical resonance at $28.2\,\mathrm{Hz}$. The SRM power reflectivity was 95\%, and the signal-recycling cavity length was $2\,\mathrm{m}$. The OPA used a periodically poled lithium niobate crystal at $33.6\,^\circ\mathrm{C}$, phase-matched for the s-polarized carrier and a 532-nm pump, and operated in a single pass; the orthogonally polarized sub-carrier was not amplified. The pump was generated using a periodically poled potassium titanyl phosphate crystal and enhanced in a bow-tie cavity of finesse 50, whose power enhancement of 25 gave an estimated $5\,\mathrm{W}$ of circulating pump power for $200\,\mathrm{mW}$ incident. The OPA acted in one propagation direction, and the same layout was used with and without pumping.

Stable operation required simultaneous control of four operating points: the interferometer at a dark fringe, the signal-recycling cavity at the desired detuning, the pump cavity on resonance, and the OPA phase for amplification of the phase signal. Unlike the gain-dependent optimization used in Fig.~\ref{fig:active-enhancement}, the amplification-axis angle was fixed at $\eta=\pi/2$. This angle amplifies the phase signal from differential mirror motion, provides a readily identifiable operating point for OPA phase control, and approaches the optimal angle in the high-gain limit. The three auxiliary fields are distinguished from the carrier by wavelength, polarization, or frequency, giving independent sensing channels for three of the four degrees of freedom, which were locked one by one without cross talk. The sub-carrier frequency offset was varied to set the signal-recycling detuning. 

In each arm the sub-carrier carried $100\,\mathrm{mW}$, the coherent-control light $30\,\mathrm{mW}$, and the 532-nm light $10\,\mathrm{mW}$. PD1 detected the 532-nm light for differential-arm-length control, with feedback to PZT1, and PD2 the p-polarized sub-carrier for signal-recycling cavity-length control, with feedback to PZT2. PD3 monitored the s-polarized dark-port output, and PD4 and PD5 detected the beats of the carrier with the sub-carrier and with the coherent-control field for their respective phase-locked loops. The coherent-control field was offset from the carrier by $31\,\mathrm{MHz}$, and its interaction with the OPA generated an idler at the opposite offset; the beat of the two in the dark-port output was demodulated at $62\,\mathrm{MHz}$ to give the OPA-phase error signal, fed back to PZT3 in the pump path.

We injected a pseudorandom white-noise excitation into PZT1, the piezo actuator driving the differential-arm length-control loop, and measured the response with an FFT analyzer. Taking the ratio of the signals recorded before and after the injection point gives the open-loop response $G_{\rm OL}(\omega)=G_{\rm SRMI}(\omega)\,G_{\rm OS,CL}(\omega)$, where $G_{\rm SRMI}$ is the passive interferometer response and $G_{\rm OS,CL}$ carries the signature of the suspended mirror.

\section{Transfer-function fitting}

\begin{figure}[htbp]
\centering
  \includegraphics[width=0.7\linewidth,trim={16 11 4 4},clip]{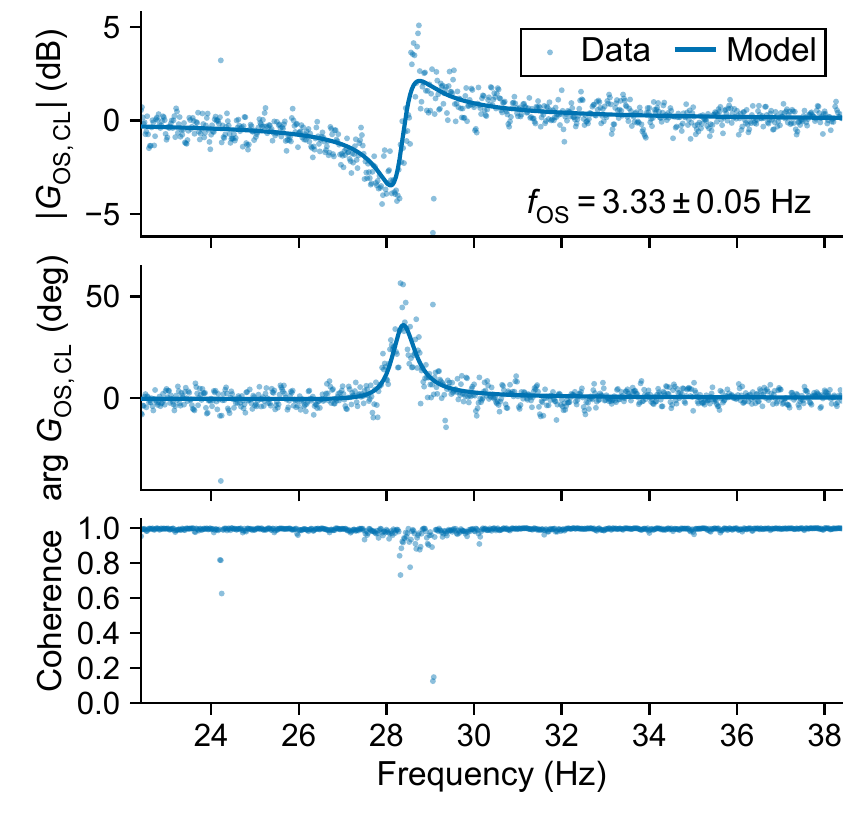}
\caption{\label{fig:results} Representative $G_{\mathrm{OS,CL}}$ measured with the OPA, with the fit giving $f_{\mathrm{OS}}=3.33\pm0.05\,\mathrm{Hz}$. The lower panel shows the coherence.}
\end{figure}

The optical spring shifts the mechanical resonance and adds damping, so $G_{\rm OS,CL}=\chi_{\rm OS}/\chi_{\rm m}$, where $\chi_{\rm m}$ is the mechanical susceptibility of the suspended mirror and $\chi_{\rm OS}$ is the mechanical susceptibility including the optical spring~\cite{Cripe2018}. The fitting model took the form
\begin{equation}
G_{\mathrm{OL}}(\omega)
=
G_{\mathrm{SRMI}}(\omega)
\frac{
\omega_{\mathrm{m}}^2-\omega^2
+2i\gamma_{\mathrm{m}}\omega
}{
\omega_{\mathrm{m}}^2+\omega_{\mathrm{OS}}^2-\omega^2
+2i(\gamma_{\mathrm{m}}+\gamma_{\mathrm{OS}})\omega
},
\end{equation}
where $G_{\rm SRMI}$ is the passive interferometer response introduced above, that is, the control-loop response excluding the optical-spring feedback, modeled as $G_{\mathrm{SRMI}}(\omega)=Ke^{i\theta_0}/(1+i\omega/\omega_{\rm c})$ with overall gain $K$, phase offset $\theta_0$ and pole frequency $\omega_{\rm c}$. The parameters $\omega_{\mathrm{m}}$, $\gamma_{\mathrm{m}}$, $\omega_{\mathrm{OS}}$, and $\gamma_{\mathrm{OS}}$ denote the mechanical resonance frequency, mechanical damping rate, optical-spring frequency, and optical damping rate, respectively.

For the complex response $H_i$ and magnitude-squared coherence $C_i$ measured at frequency $f_i$, the coherence-based weighting scale was defined as $u_i=|H_i|\sqrt{(1-C_i)/C_i}$. This scale set the relative weights of the real and imaginary residuals; its overall normalization is arbitrary and was absorbed by the rescaling of the covariance matrix described below. The real and imaginary parts were fitted simultaneously by minimizing
\begin{equation}
S
=
\sum_i
\frac{
\left[
\operatorname{Re}
\left(H_{\mathrm{model},i}-H_i\right)
\right]^2
+
\left[
\operatorname{Im}
\left(H_{\mathrm{model},i}-H_i\right)
\right]^2
}{
u_i^2
}.
\end{equation}
The formal covariance matrix was estimated using the standard linearized least-squares approximation and rescaled by the residual variance $S_{\min}/(2N-7)$, where $N$ is the number of frequency points, the factor of two accounts for the real and imaginary parts, and seven is the number of fitted parameters. Each fit provided $f_{\rm OS} = \omega_{\rm OS}/(2\pi)$, and its $1\sigma$ uncertainty $\sigma_{\rm OS}$ was obtained from the corresponding diagonal element of the rescaled covariance matrix. The optical-spring constant is proportional to $f_{\rm OS}^2$, and for the subsequent fits in $f_{\rm OS}^2$ the uncertainty $\sigma_{\rm OS}$ was transformed into asymmetric lower and upper errors,
\begin{align}
\Delta_-&=f_{\rm OS}^2-\max(f_{\rm OS}-\sigma_{\rm OS},0)^2,\\
\Delta_+&=(f_{\rm OS}+\sigma_{\rm OS})^2-f_{\rm OS}^2,
\end{align}
with $\Delta_-$ used when the model lay below the measured value and $\Delta_+$ when it lay above.

Figure~\ref{fig:results} shows a representative measurement, with the fitted $G_{\rm SRMI}$ contribution divided out to isolate $G_{\rm OS,CL}$. Repeating the fit over a range of signal-recycling detunings, with and without OPA pumping, gives $k_{\rm OS}$ as a function of detuning (Fig.~\ref{fig:results3}).

\section{Measured enhancement}

\begin{figure}[htbp]
\centering
  \includegraphics[width=\linewidth,trim={19 12 3 2},clip]{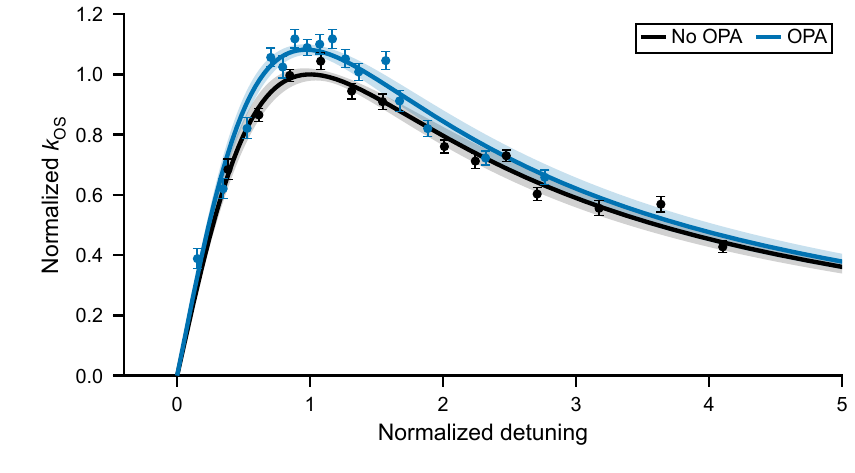}
\caption{\label{fig:results3} Optical-spring constant $k_{\mathrm{OS}}\propto f_{\mathrm{OS}}^2$ versus signal-recycling detuning, with the OPA pump off (black) and on (blue). The spring constant and detuning are normalized by the peak spring constant of the fit with the pump off and its detuning, respectively. Error bars represent the $1\sigma$ uncertainties propagated from the fitted optical-spring frequencies; shaded bands indicate the $1\sigma$ uncertainties of the fitted model curves. The fit to the OPA data includes the buildup-dependent detuning shift.}
\end{figure}

The detuning dependence in Fig.~\ref{fig:results3} was fitted with Eq.~(1) at $\eta=\pi/2$, with $r$ replaced by an effective round-trip amplitude reflectivity $r_{\rm eff}$ that includes optical losses, and with $f_0^2$ treated as a free parameter because the effective mass and power coupling of the folded arm were not determined independently. The detuning itself was not set by the sub-carrier frequency shift $f_{\rm sub}$ alone. It depended also on the shift $f_{\rm res}$ at which the carrier is resonant, which differs with and without the pump. Without the pump, $\phi=\pi(f_{\rm res}^{\rm off}-f_{\rm sub})/f_{\rm FSR}$, where $f_{\rm FSR}$ is the free spectral range of the signal-recycling cavity. Since the sub-carrier lies approximately one free spectral range below the carrier, $f_{\rm res}$ consists of $f_{\rm FSR}$ plus an offset arising from the polarization-dependent phase shifts experienced by the orthogonally polarized carrier and sub-carrier.

The pump-off data, for which $s=1$, gave $f_0^2=3.67\pm0.25$~Hz$^2$, $r_{\rm eff}=0.834\pm0.012$, $f_{\rm FSR}=75.1\pm3.6$~MHz and $f_{\rm res}^{\rm off}=87.82\pm0.12$~MHz. The first three were held fixed when fitting the pump-on data, which with $s$ and $f_{\rm res}^{\rm on}$ free gave $s=1.059\pm0.009$ and $f_{\rm res}^{\rm on}=101.71\pm0.06$~MHz.

\begin{figure}[htbp]
\centering
\includegraphics[width=0.8\columnwidth,trim={19 12 3 2},clip]{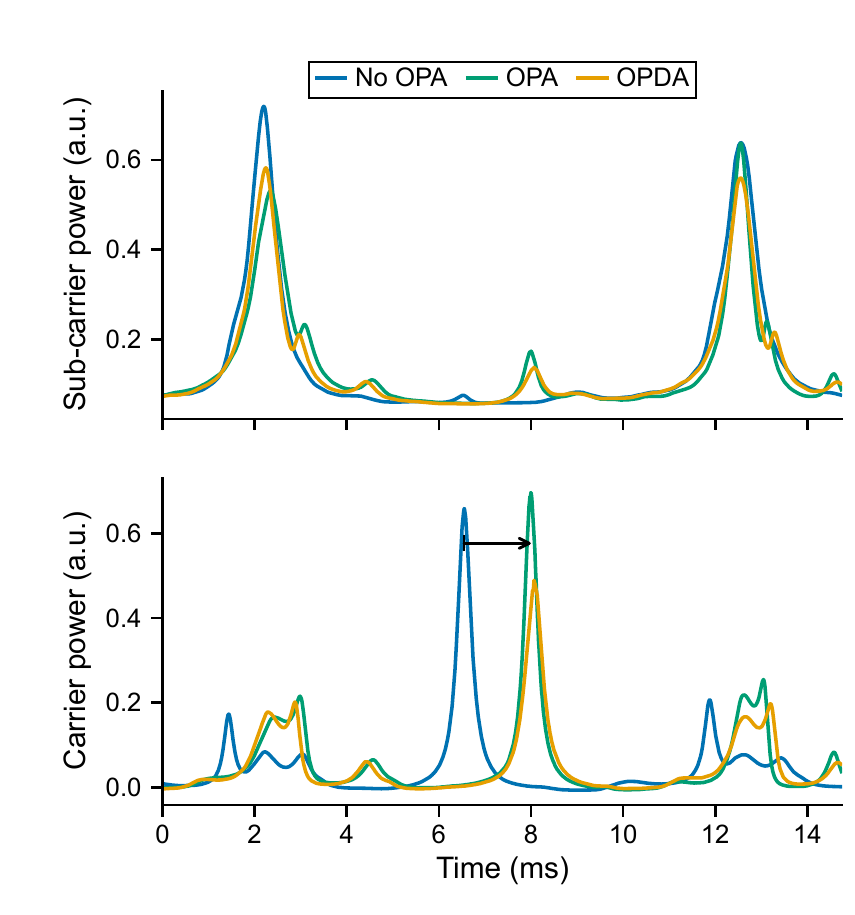}
\caption{\label{fig:results2} Pump-induced shift of the carrier resonance. Sub-carrier power at PD2 (upper) and carrier power at PD3 (lower) during signal-recycling cavity-length scans without pumping (No OPA) and at the amplification (OPA) and deamplification (OPDA) phases. A small dark-fringe offset admitted a carrier field into the cavity. The traces are aligned at the sub-carrier resonances. The arrow marks the carrier-resonance displacement.}
\end{figure}

Pump injection shifts the carrier resonance relative to the sub-carrier used for cavity control. Cavity-length scans with a small dark-fringe offset show similar shifts at the amplification and deamplification phases (Fig.~\ref{fig:results2}), indicating a pump-induced optical-path change rather than a shift specific to amplification. To include both a constant shift and a contribution dependent on intracavity power buildup, the detuning was determined self-consistently from
\begin{align}
\phi&=\pi\frac{f_{\rm res}^{\rm on}+qB(\phi)-f_{\rm sub}}{f_{\rm FSR}},\\
B(\phi)&=\frac{(1-r_{\rm eff})^2}{1+r_{\rm eff}^2-2r_{\rm eff}\cos(2\phi)},
\end{align}
where $B(0)=1$ and $q$ is the amplitude of the buildup-dependent shift, a phenomenological term that may describe additional heating from absorption of the intracavity signal field. This model gave $s=1.047\pm0.007$, $f_{\rm res}^{\rm on}=100.82\pm0.22$~MHz and $q=1.03\pm0.25$~MHz, conditional on the fixed parameters.

The reduced chi-squared was 2.6 for the pump-off data; for the pump-on data it was 7.1 without $q$ and 2.7 with it. As for the transfer-function fits, the covariance matrices were rescaled by the corresponding reduced chi-squared. The uncertainties of the fixed parameters were propagated by refitting the pump-on data with each displaced from its best-fit value and combining the resulting sensitivities with the pump-off covariance matrix, which gives $s=1.047\pm0.012$ and the shaded bands in Fig.~\ref{fig:results3}. With the enhancement defined as $\mathcal{E}=k_{\rm OS,max}^{\rm on}/k_{\rm OS,max}^{\rm off}-1$, its variance was propagated to first order as $\sigma_{\mathcal{E}}^2=(\nabla_{\bm\theta}\mathcal{E})^{\rm T}\bm{C}_{\bm\theta}(\nabla_{\bm\theta}\mathcal{E})$, where $\bm\theta=(r_{\rm eff},s)^{\rm T}$ and $\bm{C}_{\bm\theta}$ is the joint covariance matrix of the two fits. The fitted peak ratio was $1.082\pm0.028$, an enhancement of $8.2\pm2.8\%$.

Comparing the largest measured optical-spring constants directly, with their uncertainties treated as uncorrelated, gives a ratio of $1.071\pm0.040$, or $7.1\pm4.0\%$. This avoids the detuning model but uses the same data without correcting for the pump-induced shift, so it is a cruder estimate rather than an independent determination.

\section{Summary and discussion}\label{sec:discussion}

We implemented a parametric amplifier in the signal-recycling cavity of a Michelson interferometer operated at a dark fringe, and observed an increase of $8.2 \pm 2.8\%$ in the peak optical-spring constant. The inferred OPA gain is $s = 1.047 \pm 0.012$. Separating the signal field from the carrier in this way provides a simple and effective means of manipulating the optical spring without further increasing the carrier power, and without exposing the nonlinear crystal to the circulating carrier field. The main systematic limitation of the present apparatus is the pump-induced shift of the carrier resonance relative to the sub-carrier used for cavity control, which required the detuning to be inferred from a model rather than read from the sub-carrier offset alone.

At zero detuning, Eq.~(2) places the parametric oscillation threshold at $s_{\rm th}=1.199\pm0.017$ for the measured $r_{\rm eff}=0.834$, which corresponds to a single-pass power loss of 14\%. The achieved gain, $s=1.047\pm0.012$, is $0.40\,\mathrm{dB}$, a factor of four below the $1.58\,\mathrm{dB}$ at threshold. Since $\ln s$ scales as the square root of the pump power, reaching the threshold with the present optics would require $3.1\,\mathrm{W}$ rather than the $200\,\mathrm{mW}$ used here; with the present pump it is reached once the single-pass loss is brought down to about 2\%. Intermediate combinations are available, halving the loss bringing the required pump to $1.0\,\mathrm{W}$. The route to a strong optical spring is therefore a specified trade between intracavity loss and pump power, neither of which involves increasing the carrier power.

Further enhancement is therefore limited by optical loss and by the operating range of the intracavity amplifier rather than by carrier-power handling. Active elements in the signal-recycling cavity have been proposed as a means of tailoring the signal response and quantum noise of interferometric sensors~\cite{Miao2015,Bentley2019,Korobko2019,Vermeulen2026,Suzuki2026}; the present experiment realizes one and measures its effect on the interferometer response.



\begin{acknowledgments}
This work was supported by the Japan Society for the Promotion of Science (23KJ0954, 17H02886, and 23K25896) and the Japan Science and Technology Agency (JPMJSC2209, JPMJCR1873, and JPMJAP2320).
\end{acknowledgments}



\bibliography{references}

\end{document}